\documentclass[12pt]{elsarticle}

\usepackage{amssymb}
\usepackage{amsthm}
\usepackage{float}
\usepackage[left=2.5cm, right=2.5cm, top=2.5cm, bottom=2.5cm, footskip=0.5cm]{geometry}

\usepackage[usenames,dvipsnames]{color} %Color package

\journal{ }

\usepackage{subfig}

\begin{document}

\begin{frontmatter}

%\noindent
%{\color{RubineRed} \rule{\linewidth}{0.5mm}}

\title{\textbf{Numerical Study of the Wetting Dynamics of Droplet on Laser Textured Surfaces: Beyond Classical Wenzel and Cassie-Baxter Model}}
\date{ }

\author[1]{Ilemona S. Omeje}
\author[1]{Tatiana E. Itina}
%\author[1]{Author Three}

\address[1]{Laboratoire Hubert Curien, UMR CNRS 5516, Université Jean Monnet, Universite de Lyon, Saint-Etienne, France\\
Corresponding author: ilemona.sunday.omeje@univ-st-etienne.fr}
%\address[2]{School of Science, University of Technology, 2000 J St. NW, Washington DC, 20036, USA}
%\address[3]{Center Of Intelligence, Faculty of Sciences, 50 West 4th Street, New York NY, 37895, USA}
%\address[1]{Corresponding author: ilemona.sunday.omeje@univ-st-etienne.fr}

 %\vspace{0.5cm}

\begin{abstract}


The classical wetting models, such as the Wenzel and Cassie-Baxter have been extensively used to quantify the wettability of laser-textured surfaces. However, these models do not provide any description of the corresponding droplet dynamics. In this work, we propose a detailed continuum-level modelling to study the wetting dynamics of a water droplet on Ti-6Al-4V alloy. The calculations are performed for flat surfaces and surfaces with various reliefs. The calculated evolutions of the droplet spreading parameter for flat surfaces, surfaces with triangular reliefs and one with two different periods and heights not only provide explanations of several experimental results but also underline the perspectives of using complex reliefs for efficient wettability control. Thus, such simulations are shown to be useful in relief design and laser texturing for a wide range of applications, for example, laser treatment of artificial implants and prostheses.

\end{abstract}


\end{frontmatter}

\section{Introduction}
\label{sec:sample1}

%%       \citet{<label>} ==> Jones et al. [21]
%%       \citep{<label>} ==> [21]
%%

It is well-known that laser pulses interacting with solid surfaces often create a set of periodic or non-periodic surface structures ranging from "ripples" also known as laser-induced periodic surface structures (LIPSS) to grooves and spikes. Additionally, when laser scanning systems are used, so-called hierarchical structures can be easily formed on surfaces combining both micro- and nano-reliefs and allowing a wide-range control over surface wettability  \citep{doi:10.2351/1.4712658}\citep{doi:10.1021/la8037582}\citep{BIZIBANDOKI2013399}. One of the promising applications of such effects is in the integration and durability of dental and orthopedic implants \citep{doi:10.1098/rsta.2010.0003}\citep{LIU200449}. Other modern applications include self-cleaning glasses and metals, as well as microfluidic effects in-jet printing \citep{Barthlott1997}\citep{https://doi.org/10.1002/lpor.201200017}\citep{wu2021robust}. 
When femtosecond lasers are used, the control possibilities are considerably enlarged. Depending on laser parameters, femtosecond textured designed surfaces can be either high-spatial frequency or low-frequency LIPPSS with various orientations, cones, pillars, as well as various combinations of these structures that can be obtained by using additional optical filters and polarization devices, laser beam interferometry set-ups, double-pulse irradiation, etc.  \citep{doi.org/10.1002/admi.201701370}.  

The wettability of laser textured surfaces changes not only as a function of surface relief, but also with time \citep{BIZIBANDOKI2013399}\citep{CUNHA2013688}. The reasons for these changes are currently under investigation. Several time scales have been revealed, the shorter one being related to the droplet evolution upon its interaction with the surface, while the longer one is often related with the changes in surface chemistry due to oxidation, molecular adsorption, etc. Extensive experimental work has been performed to examine the influence of femtosecond textured designed surfaces on wettability to improve implant conditions. For example, Cunha et. al. \citep{CUNHA2013688} reported that biomedical grade titanium-6aluminum-4vanadium (Ti-6Al-4V) alloy surfaces textured by femtosecond laser possess anisotropic wetting properties. Klos et. al. \citep{nano10050864} further reported that laser-induced topographies on Ti-6Al-4V alloy influenced surface wettability and protein adsorption with promotion of osteogenic differentiation. 

The wettability of an ideal homogeneous and smooth surface is commonly described by the classical Young’s law \citep{doi:10.1098/rspl.1800.0095} that balances the three interfacial forces at the contact-line of solid-vapor, solid-liquid and liquid-vapor. Young’s equation predicts the intrinsic equilibrium contact angle (CA), which relates to the surface tension between the three phases. In fact, surface roughness is an important parameter controlling the contact angle and hence, the wettability \citep{doi:10.1021/j150474a015}\citep{doi:10.1021/ie50320a024}. Wenzel \citep{doi:10.1021/ie50320a024} and Cassie-Baxter \citep{TF9444000546} proposed modified theoretical models to correct Young’s equation by accounting for surface roughness and the presence of air entrapped on textured surfaces. In the past decades, the advancing and receding contact angles were replaced by an equilibrium contact angle in the corrected models to interpret the behavior of liquid on rough surfaces. Furthermore, Wenzel model predicts that roughness enhances the intrinsic wetting behavior of a surface with the assumption that liquid fully penetrates the solid roughness with a roughness factor. Wenzel argument was partially validated experimentally by Onda et. al. \citep{doi:10.1021/la950418o} and Shibuichi et. al. \citep{doi:10.1021/jp9616728}. However, Onda et. al. \citep{doi:10.1021/la950418o} revealed the limitations of Wenzel model for extremely rough surfaces. On the other hand, Cassie-Baxter model \citep{TF9444000546} assumed that, on such surface, a large fraction of air is entrapped so that the liquid do not completely penetrate the textured surface. Liquid invasion on the textured surface takes place through capillary phenomenon, while the remaining liquid stays on the surface \citep{doi:10.1021/la1039893}. Quere \citep{2005} and Bico et. al. \citep{BICO200241} attempted to combine the two models to explain the effect of roughness on wetting state. In addition, it is unclear whether Wenzel and Cassie-Baxter model can accurately predict some of the experimental trends of laser textured surfaces. In fact, McHale \citep{doi:10.1021/la7011167} reported that Wenzel and Cassie-Baxter equations can be applied only (a) when the surface is similar and isotropic and (b) when the droplet size ratio to the roughness is large enough. In most cases, these models are often oversimplified for wetting studies of laser textured surfaces. The development of much more realistic models is challenging but it is required for better relief design, namely in many modern laser applications.  

Several numerical models have been proposed and deployed to study the wettability of a surface, wetting interaction at the three-phase (i.e solid-liquid-vapor interface) and droplet spreading mechanism. For example, Sikalo et.al. \citep{doi:10.1063/1.1928828} used the Volume of Fluid (VOF) method to study the effect of a droplet on a flat surface. Meanwhile, Grewal et. al. \citep{C4NR04069D} used the Level Set Method (LSM) to show that topography along with the surface chemistry and geometrical parameters control the wetting performance of patterned surfaces. Chamakos et. al. \citep {doi:10.1063/1.4941577} tried to tackle the contact-line boundary condition of spreading dynamics on textured substrates by using the continuum-level approach. Also, Yagub et. al. \citep{YAGUB2015402} used Shan and Chen (SC) lattice Boltzmann model (LBM) \citep{PhysRevE.47.1815} to highlight the strength and weaknesses of the SC model in the description of changes in the droplet apparent contact angle observed on nano/microstructured surfaces. 
Despite an enormous interest and a high number of studies, many questions have not been enough addressed yet. For instance, the effects of complex or hierarchical surface reliefs were not sufficiently studied. There is still a lack of understanding of the dynamics of liquid droplets on such surfaces. 

In this work, we use a continuum-level modelling method such as LSM to model the wetting dynamics of a water droplet on Ti-6Al-4V alloy. The wetting studies are performed on flat surfaces, surfaces with triangular reliefs and one with two different periods and heights. The spreading parameter (or diameter) of the droplet as a function of the contact time is used to examine the short-scale droplet dynamics on several textured materials. In particular, we considered the effects of surface reliefs, velocity and viscosity of droplet evolution on Ti-6Al-4V alloy surfaces.

\section{Modeling details}
\label{sec:another}

Firstly, we started by developing a model of a two-phase flow on a flat surface.
For simplicity, the following assumptions have been made. (a) Two immiscible fluids, air and liquid (e.g. water), on plane (and relief) surfaces are assumed to be Newtonian, incompressible and laminar. (b) The droplet-wall collision is considered to be isothermal i.e., both the fluid and the surface are at constant temperature and (c) no heat transfer is considered during the interaction between the wall and the liquid (i.e. adiabatic case). The two-phase flow dynamic is simulated by using the time-dependent incompressible Navier-Stokes and continuity equations for conservation of mass and momentum as shown in eqs. 1 and 2. 

\newcommand{\dd}[1]{\mathrm{d}#1}

\begin{equation}\label{equation_one}
\rho(\frac{\partial \mathbf{u}}{\partial t}+(\mathbf{u}.\nabla){\mathbf{u}}) = \nabla.[-p\mathbf{I} + \mu(\nabla \mathbf{u} +(\nabla \mathbf{u})^T) + \rho.\mathbf{g} + \mathbf{F_{s_t}} ]
\end{equation}
\begin{equation}\label{equation_two}
\rho\nabla.\mathbf{u} = 0
\end{equation}

Here $\rho$ and $\mathbf{u}$  are the fluid density ($kg/m^3$) and velocity vector (m/s); p denotes the pressure (Pa), $\mathbf{I}$ is the second-order identity matrix, $\mu$ represents the dynamic viscosity (Pa.s), $\mathbf{g}$ is the gravitational acceleration ($m/s^2$) and $\mathbf{F_{s_t}}$  represents the surface tension force (N/m). The surface tension force is calculated as follows.
\begin{equation}\label{equation_four}
\mathbf{F_{s_t}} = \nabla.[(\sigma(\mathbf{I}-\mathbf{nn}^T))\delta]
\end{equation}
$\mathbf{I}$  is the identity matrix, $\mathbf{n}$ is the normal to the interface, $\sigma$ is the surface tension and $\delta$ is Dirac delta function that is nonzero only at the fluid interface. Because the two fluids are immiscible, the conservative level set method (LSM) proposed by Olsson and Kreiss \citep{OLSSON2005225} is used to track the fluid interface. The interface is captured by a level-set function represented by a smeared Heaviside function ($\phi$). $\phi$ changes smoothly across the interface from 0 (in air) to 1 (in liquid) and the interface is represented by $\phi$ = 0.5 (Fig. 1). The LSM eqs. is shown in eqs. (4).

\begin{equation}\label{equation_three}
\frac{\partial\phi}{\partial t} + \mathbf{u}.\nabla\phi = \gamma\nabla.(\varepsilon_l\nabla\phi - \phi(1-\phi)\frac{\nabla\phi}{|\nabla\phi|}) 
\end{equation}

Here, $\varepsilon_l$ and $\gamma$ represents the artificial thickness of the interface and the re-initialization parameter of the level set function. $\varepsilon_l$ value is taken as half of the mesh size and $\gamma$ value is defined as the initial value of impact velocity of droplet \citep{hu2014simulation}\citep{2018}. The right and left-hand sides of eqs. 4 are necessary for numerical stability and movement of the interface. Both $\rho$ and $\mu$ are functions of $\phi$ as shown in eqs. 5 and 6, and are considered to vary smoothly over the interface.

\begin{equation}\label{equation_three}
\rho = \rho_1 + (\rho_2 +\rho_1)\phi
\end{equation}
\begin{equation}\label{equation_three}
\mu = \mu_1 + (\mu_2 +\mu_1)\phi
\end{equation}

Here, $\rho_1$, $\rho_2$, $\mu_1$  and $\mu_2$ are the densities and dynamic viscosities of air and liquid (e.g. water). Eqs. 1 to 6 are calculated by using the finite-element solver discretized in COMSOL Multiphysics 5.6 software \citep{2018}. The simulation is performed in 2D axis-symmetry (Fig. 1). Open boundary conditions were imposed at all the boundaries with zero pressure except for the wetted wall. In addition, equilibrium contact angle (ECA) was used as the solid boundary condition at the triple contact line for the setting of the surface wettability instead of static contact angle. The ECA is estimated from the advancing ($\theta_{adv}$) and receding CA ($\theta_{rec}$) by using the theoretical work of Tadmor \citep{doi:10.1021/la049410h} as follows. 

\begin{equation}\label{equation_three}
\theta_{ECA} = cos^{-1}(\frac{r_{adv}cos(\theta_{adv})+r_{rec}cos(\theta_{rec})}{r_{adv}+r_{rec}})
\end{equation}

where,

\begin{equation}\label{equation_three}
r_{adv} = (\frac{sin^3(\theta_{adv})}{2-3cos(\theta_{adv})+cos^3(\theta_{adv})})
\end{equation}

and,

\begin{equation}\label{equation_three}
r_{rec} = (\frac{sin^3(\theta_{rec})}{2-3cos(\theta_{rec})+cos^3(\theta_{rec})})
\end{equation}

Navier slip boundary condition is adopted for the wetted wall and the contact angle to avoid singularity. Thus, a frictional force ($\mathbf{F_{fr}}$ = -{$\mu_\phi$}$\mathbf{u}$/{$\Gamma$}) is added on the liquid-solid interface, where $\Gamma $ is the slip length, which is taken as the default size of the mesh \cite{2018} and $\mathbf{u}$ is the slip length velocity. 

\begin{figure}[h]
    \centering
    \includegraphics[width=0.45\textwidth]{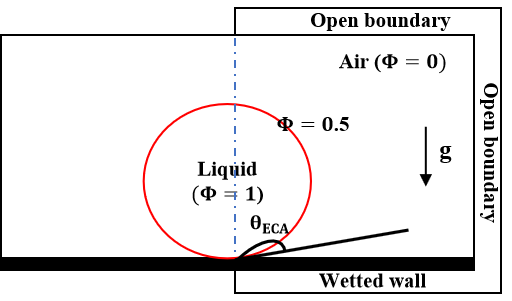}
    \caption{2D axis-symmetry simulation domain.}
    \label{fig:First}
\end{figure}

\subsection{Model validation}

To check the validity of the model described above, we performed a series of calculations for water droplet impact on surfaces and compared the results obtained with the experimental results of Lin et. al. \citep{LIN201886} and Yokoi et.al. \citep{doi:10.1063/1.3158468}.  
The model was then adopted for triangular reliefs and for one with two different periods and heights in order to elucidate the optimum relief required to understand the wetting dynamics of laser textured materials for biomedical applications.

\begin{table}[ht]
\caption{Estimated equilibrium contact angle ($\theta_{ECA}$)} % title of Table
\centering % used for centering table
\begin{tabular}{c c c c} % centered columns (4 columns)
\hline\hline %inserts double horizontal lines
Type of plane surface & Advancing CA ($\theta_{adv}$)  & Receding CA ($\theta_{rec}$) & Estimated $\theta_{ECA}$ \citep{doi:10.1021/la049410h} \\ [1ex] % inserts table
%heading
\hline % inserts single horizontal line
N-BK7 glass \citep{LIN201886} & 111$^o$ & 100$^o$ & 106$^o$ \\ % inserting body of the table
Dry substrate \citep{doi:10.1063/1.3158468} & 105$^o$ & 75$^o$ & 90$^o$ \\ [1ex] % [1ex] adds vertical space
\hline %inserts single line
\end{tabular}
\label{table:nonlin} % is used to refer this table in the text
\end{table}

The first case of validation was performed by using the experimental data of Lin et. al. \citep{LIN201886}. The diameter of water droplet ($D_o$) on N-BK7 glass is 2 mm with an initial impact velocity ($u_o$) of 0.52 m/s and dynamic viscosity ($\mu$) of 0.9 mPa.s at a surface tension ($\sigma$) of 71.8 mN/m. The density ($\rho$) of the droplet is 997 $kg/m^3$ and the estimated ECA ($\theta_{ECA}$) used in the simulation is 106$^o$ as shown in Table 1. The results of the 2D and 3D time-lapse images of water droplet impact on the N-BK7 glass surface and those of the current study are presented in Fig. 2. Excellent agreement is obtained with the experimental results of Lin et. al. \citep{LIN201886} and our numerical simulation. Before the impact, the droplet shape is spherical, afterward, the droplet is deformed and spreads rapidly on the surface. It starts to recoil at 3 ms. Interestingly, at 7.25 ms, entrapped bubble is observed. This effect is due to the fast rate of recoiling and a capillary wave formation \citep{doi:10.1063/1.1928828}\citep{doi:10.1063/1.1527044}.

\begin{figure}[h]
    \centering
    \includegraphics[width=1\textwidth]{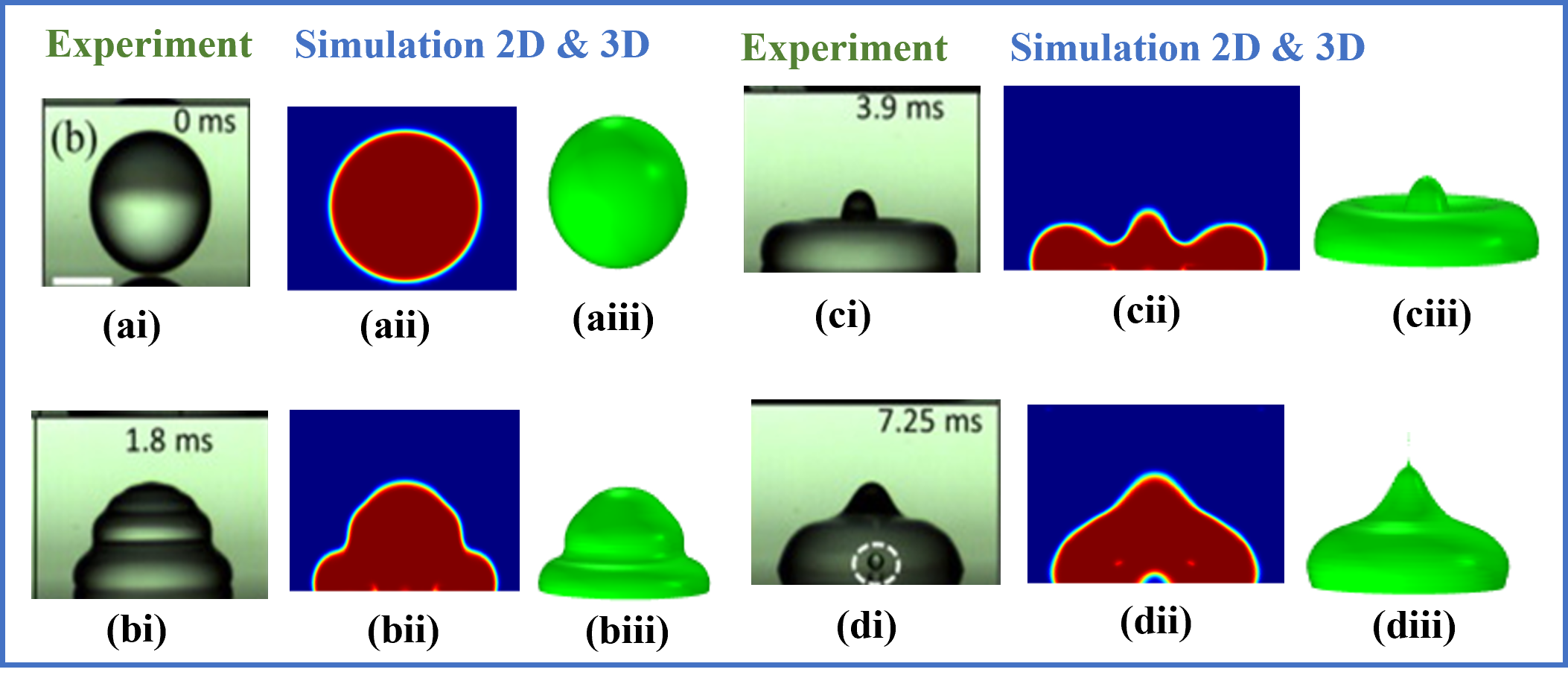}
    \caption{Comparison of experimental data of \citep{LIN201886} with 2D and 3D droplet of the current study at 0, 1.8, 3.9 and 7.25 ms respectively.}
    \label{fig:First}
\end{figure}

%\begin{figure}[h]
%    \centering
%    \includegraphics[width=0.8\textwidth]{Result_Original.png}
%    \caption{Spreading parameter of water on wax surface.}
%    \label{fig:First}
%\end{figure}
%\begin{figure}[h]
%    \centering
%    \includegraphics[width=0.8\textwidth]{Mesh_studies.png}
%    \caption{Grid convergence test.}
 %   \label{fig:First}
%\end{figure}

%\usepackage{subfigure}

\begin{figure}%
    \centering
    \subfloat{{\includegraphics[scale=0.506]{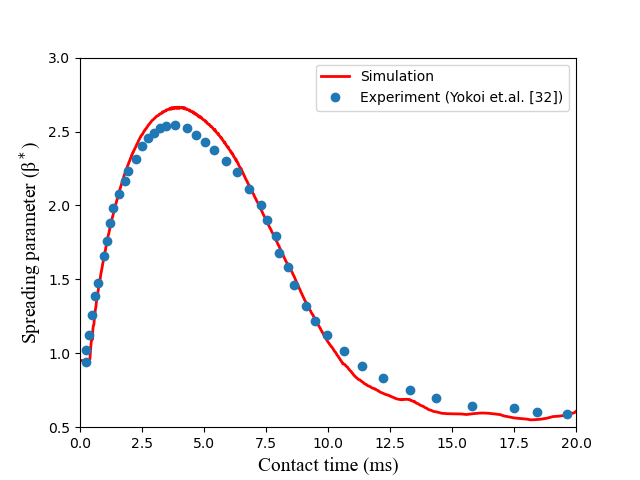} }}%
    %\qquad
    \subfloat{{\includegraphics[scale=0.506]{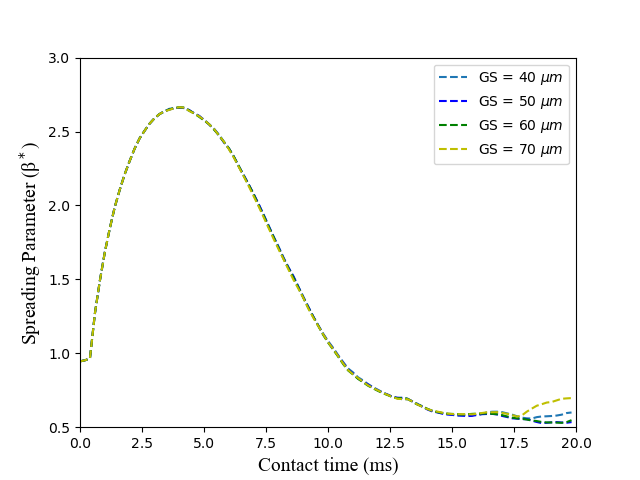} }}%
    \caption{(a) Spreading parameter ($\beta^*$) comparison of \citep{doi:10.1063/1.3158468} with current simulation (b) Mesh convergence test.}%
    \label{fig:example}%
\end{figure}

%\begin{figure}
%    \centering
%    \subfigure{\includegraphics[scale=0.506]{sande.png}} 
%    \subfigure{\includegraphics[scale=0.506]{gtest.png}} 
%    %\subfigure[]{\includegraphics[width=0.24\textwidth]{monalisa.jpg}}
%    %\subfigure[]{\includegraphics[width=0.24\textwidth]{monalisa.jpg}}
%    \caption{(a) Spreading parameter of water on wax surface (b) Grid convergence test. }
%    \label{fig:foobar}
%\end{figure}

The second model validation was carried out by using the experimental study of Yokoi et.al. \citep{doi:10.1063/1.3158468}. The initial droplet diameter ($D_o$) of water is 2.28 mm with an impact velocity ($u_o$) of 1 m/s at estimated ECA ($\theta_{ECA}$) of 90$^o$. The surface tension ($\sigma$) is 7.2 mN/m, dynamic viscosity ($\mu$) and density ($\rho$) of the droplet is 1 mPa.s and 1000 $kg/m^3$. The plot of the spreading parameter ($\beta^*$) of a water droplet on a dry substrate is compared with the experimental findings of \citep{doi:10.1063/1.3158468}. It should be noted that the spreading parameter $(\beta^*)$ is calculated by dividing the droplet wet diameter (d) by the initial droplet diameter ($D_o$). 
As presented in Fig. 3(a),  the calculated $\beta^*$ agrees well with the experimental values. Moreover, during both the impact and the recoil of the water droplet on the dry substrate, an excellent agreement is observed. However, a small deviation can be noticed at the peak of the $\beta^*$. We believe the noticed small difference in the curves in Fig. 3(a) is within the experimental errors. In fact, according to the experiments of Yokoi et.al. \citep{doi:10.1063/1.3158468}, the receding contact angle is 75$^o$ ($\pm$2$^o$) and the advancing contact angle is 105$^o$ ($\pm$2$^o$). In the calculations, we used 75$^o$ and 105$^o$ as the receding and advancing contact angle as shown in Table 1.

Before the simulation of water droplet on different Ti-6Al-4V relief surfaces, we performed a mesh sensitivity test of water droplet on dry substrate to see where the numerical simulation result converges. For this, different grid sizes (GS) from 40 $\mu m$ to 70 $\mu m$ with a step size of 1mm was used to calculate $\beta^*$ as presented in Fig. 3(b). It is observed that the simulation results of $\beta^*$ converges at a grid size of 40 $\mu m$ to 60 $\mu m$ with little deviation at 70 $\mu m$. We, therefore, maintained a grid size of 40 $\mu m$ throughout this study. Again, the results of the numerical model and the experimental results match. It is also worth noting that the water droplet is sensitive to the substrate surface \citep{https://doi.org/10.1002/aic.690430903}. Other liquid interactions on flat surfaces can be calculated with required parameters, such as surface tension, dynamic viscosity, and density. So, the model  predicts accurately enough the droplet interaction and spreading on several surfaces.

\section{Result and discussion}
\label{sec:another}    
\subsection{Effects of surface reliefs on wettability of several textured Ti-6Al-4V surfaces}
The periodic patterns formed by femtosecond laser irradiation are known to be of several types: (i) below the wavelength, in the range of a few tens of nanometers (high spatial frequency LIPSS, or HSFL); (ii) low spatial frequency LIPSS (LSFL or, grooves), and (iii) complex 3D microstructures with several feature sizes and periods \citep{nano11040899}\citep{varlamova2019wetting}\citep{dominic2021insignificant}. Herein, we consider a triangular relief with one (Fig. 4(a)) and with two periods and heights (Fig. (b)) on Ti-6Al-4V alloy. This allows us to mimic, at the first approximation, the typical triangular pattern of laser-textured surfaces \citep{nano11040899}\citep{liu2021femtosecond}. To capture the experimental trends observed for sub-micrometer LIPPS, we considered a droplet diameter larger than the corresponding structural reliefs. The area fractions ($\phi_s$ = 0.5pq) of the flat surface, triangular and two-period reliefs, that are assumed to be covered by air are 0 $\mu m^2$, 0.375 $\mu m^2$ and 1.1 $\mu m^2$ respectively. Here, p is the height and q is the width of the structure. A typical representation of these reliefs is presented in Fig. 4(a) and 4(b).

\begin{figure}[h]
    \centering
    \includegraphics[width=1\textwidth]{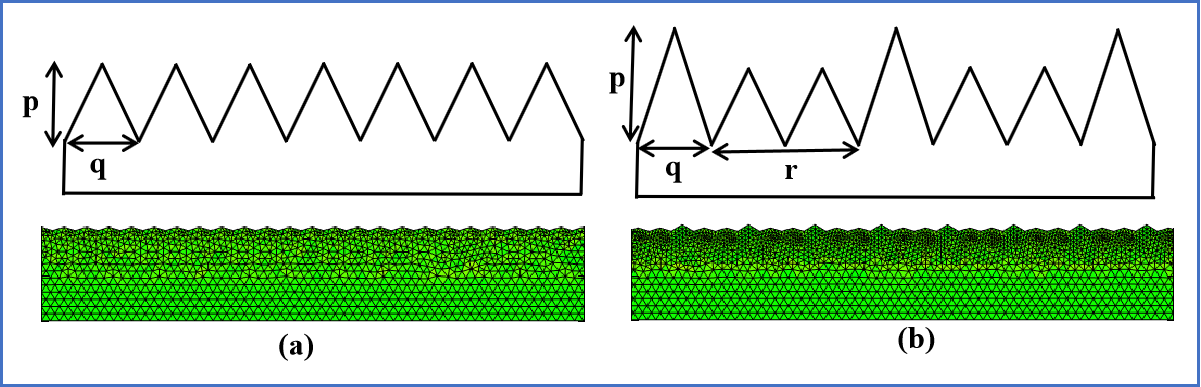}
    \caption{(a) Triangular and (b) a two-period reliefs with different heights used in our calculation.}
    \label{fig:First}
\end{figure}

%\begin{figure}
%    \centering
%    \subfigure{\includegraphics[scale=0.8]{ph.png}} 
%    %\subfigure{\includegraphics[scale=0.7]{Second_R_2.png}} 
%    %\subfigure[]{\includegraphics[width=0.24\textwidth]{monalisa.jpg}}
%    %\subfigure[]{\includegraphics[width=0.24\textwidth]{monalisa.jpg}}
%    \caption{(a) Periodic triangular and (b) hierarchical surface reliefs used in our simulations. }
%    \label{fig:foobar}
%\end{figure}

The diameter ($D_o$) of water droplet is 2.35 mm with an impact velocity ($u_o$) of 1 m/s at an equilibrium contact angle ($\theta_{ECA}$) of 96$^o$. The Reynolds number, Re = $\rho$$u_o$$D_o$/$\mu$ is 2345 and Weber number, We = $\rho$$u^2$$D_o$/$\sigma$ is 32. Where $\rho$ is 998 kg/m$^3$, $ \sigma$ is 73 mN/m and $\mu$ is 1 mPa.s. The deformation of water droplet on flat, triangular and two-period Ti-6Al-4V reliefs are shown in Fig. 6 (a), (b) and (c). The calculated spreading diameter ($\beta^*$) and droplet thickness ($\alpha^*$ = h/$D_o$) of water droplet on Ti-6Al-4V relief surfaces are illustrated in Fig. 5 (a) and (b). Here, h is the droplet height. The hydrodynamic of the droplet impact on flat (Ti-6Al-4V) surfaces are well known\citep{doi:10.1063/1.1527044}\citep{https://doi.org/10.1002/aic.690430903}. The droplet deformation and the observed shape with the difference in the central part and the periphery are often observed on surfaces \citep{https://doi.org/10.1002/aic.690430903}. As shown in Fig. 6(a), the droplet is spherical at 0 s prior to the impact but starts to spread with an observed rim formation at 800 $\mu s$. It reaches a maximum $\beta^*$ of 2.6 mm (Fig. 5(a)) at 4 ms. The rim at the periphery thickens at 4ms due to viscosity and surface tension \citep{doi:10.1063/1.1527044}\citep{thoroddsen2005air}. The droplet starts to retract at 6.6 ms and at 9 ms, the droplet rebounds. In this study, we only considered the spreading phase of the water droplet. Similarly, the droplet has a thickness of 2.45 mm at the initial stage, see Fig. 5 (b). As it deforms to reach a maximum spreading, the thickness of the droplet is 0.71 mm at 3.4 ms before it starts to recoil. 

\begin{figure}%
    \centering
    \subfloat{{\includegraphics[scale=0.506]{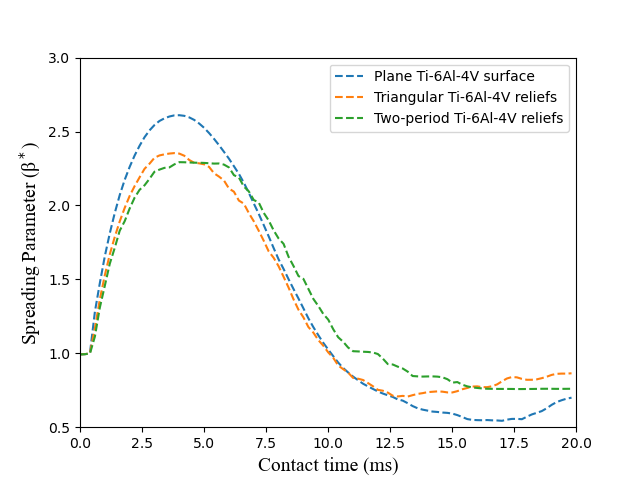}}}%
    %\qquad
    \subfloat{{\includegraphics[scale=0.506]{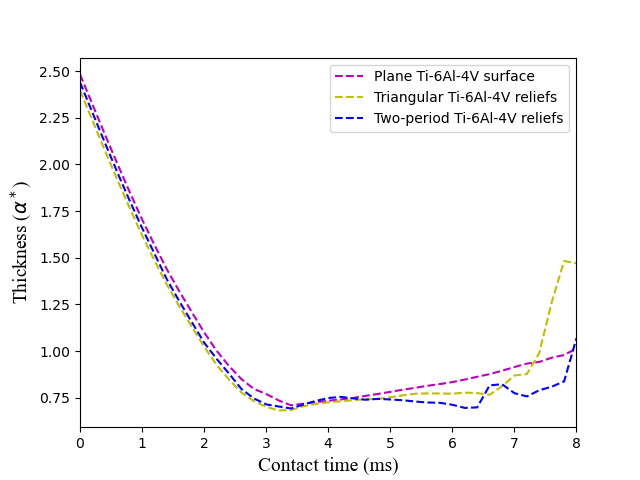} }}%
    \caption{(a) Spreading parameter ($\beta^*$) and (b) thickness ($\alpha^*$) of water droplet on plain, triangular and a two-period reliefs Ti-6Al-4V alloy surface. }%
    \label{fig:example}%
\end{figure}

%\begin{figure}
%    \centering
%    \subfigure{\includegraphics[scale=0.506]{pth.png}} 
%    \subfigure{\includegraphics[scale=0.506]{tpth.png}} 
%    %\subfigure[]{\includegraphics[width=0.24\textwidth]{monalisa.jpg}}
%    %\subfigure[]{\includegraphics[width=0.24\textwidth]{monalisa.jpg}}
%    \caption{(a) Spreading parameter ($\beta^*$) and (b) thickness ($\alpha^*$) of water droplet on plain, triangular and hierarchical Ti-6Al-4V alloy surface. }
%    \label{fig:foobar}
%\end{figure}

\begin{figure}[h]
    \centering
    \includegraphics[width=1\textwidth]{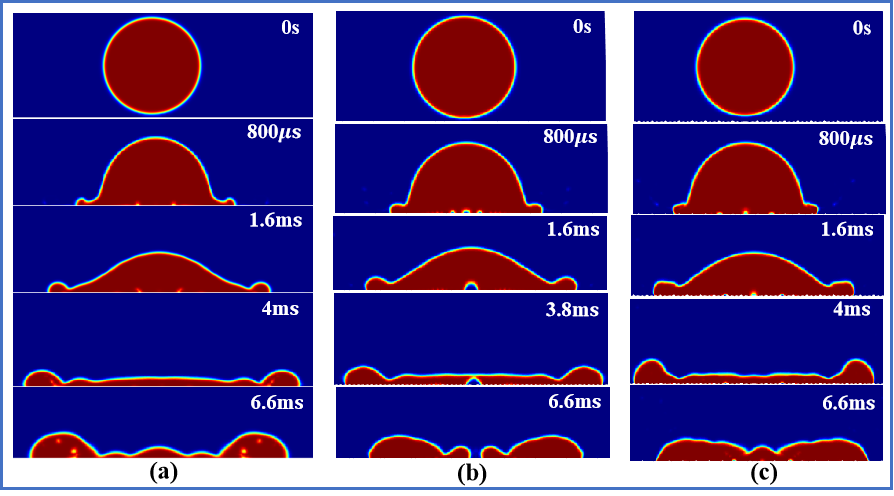}
    \caption{Snapshot of water impact on (a) solid, (b) triangular, and (c) two-period reliefs Ti-6Al-4V alloy surface with $u_o$ of 1m/s at $\theta_{ECA}$ of 96$^o$.}
    \label{fig:First}
\end{figure}

When a triangular-like relief with surface fraction ($\phi_s$) of 0.375 $\mu m^2$ was introduced on Ti-6Al-4V surface, as presented in Fig. 4(a), the evolution of the water droplet and wettability changes significantly (Fig. 6(b)). In fact, when the droplet starts to spread at 800 $\mu s$, an air bubble is formed in the center of the droplet. The air bubble formation at the solid-liquid interface is due to the air pressure in the gap between the reliefs. And as the water droplet reaches a maximum spreading of 2.62 at 3.8 ms on the triangular relief, air is forced out because of the pressure difference below the water droplet on the introduced reliefs \citep{doi:10.1063/1.1527044}\citep{thoroddsen2005air} (Fig. 6(b)). The $\phi_s$ of flat Ti-6Al-4V surface covered by the water droplet is zero as it is assumed to be smooth and the calculated $\theta_{ECA}$ at maximum spreading diameter is 102$^o$, Fig. 7(bI). However, when triangular relief was introduced, it reaches the maximum spreading diameter of 2.62 at 3.8 ms with $\phi_s$ of 9.9 $\mu m^2$ at an increased $\theta_{ECA}$ of 120$^o$ (Fig. 7(bII)) contrary to that on the smooth surface. In addition, when the height (p) of the triangular relief was increased (in-between two-period reliefs) as shown in Fig.4(b), the maximum spreading diameter observed is 2.25. Here, $\phi_s$ covered by water is 10.7 $\mu m^2$ at the calculated $\theta_{ECA}$ of 122$^o$ as presented in Fig. 7(bIII). The $\phi_s$ for the two introduced reliefs shows that the spreading parameter becomes weaker as the roughness scale increases, Fig. 5(a). Nonetheless, the reliefs still favor a rim-like structure formation at the periphery of the droplet with an increase in contact angle of the droplet. The obtained results show, therefore, how the considered small sub-micrometer surface reliefs affect the droplet speeding. Particularly, the considered triangular relief with one period (Fig. 4(a)) changes the wetting dynamics and increases the spreading parameter, while the one with two periods and heights (Fig. 4(b)) confines the droplet and decreases its spreading. There results indicate, namely, that typical HSFL alone are probably not well suitable for the increase in hydrophobic properties of the considerable surfaces, but rather two-period or more complex combinations of LSFL with HSFL systems are expected to be much more promising for this. We note that the possibility of formation of such combinations is one of the main advantages of femtosecond laser systems, so the perspectives of their applications for wettability applications is clearly confirmed by our simulations. The calculated $\theta_{ECA}$ at maximum $\beta^*$ for the three surfaces as shown in Fig. 7 (bI, bII and bIII) are 102$^o$, 120$^o$ and 122$^o$ respectively. The measured $\theta_{ECA}$ is used in the next section to validate the maximum spreading diameter based on energy analysis. It should be noted that the measured $\theta_{ECA}$ was calculated by using open-source software \citep{Schneider2012}. 

\begin{figure}%
    \centering
    \subfloat{{\includegraphics[scale=0.68]{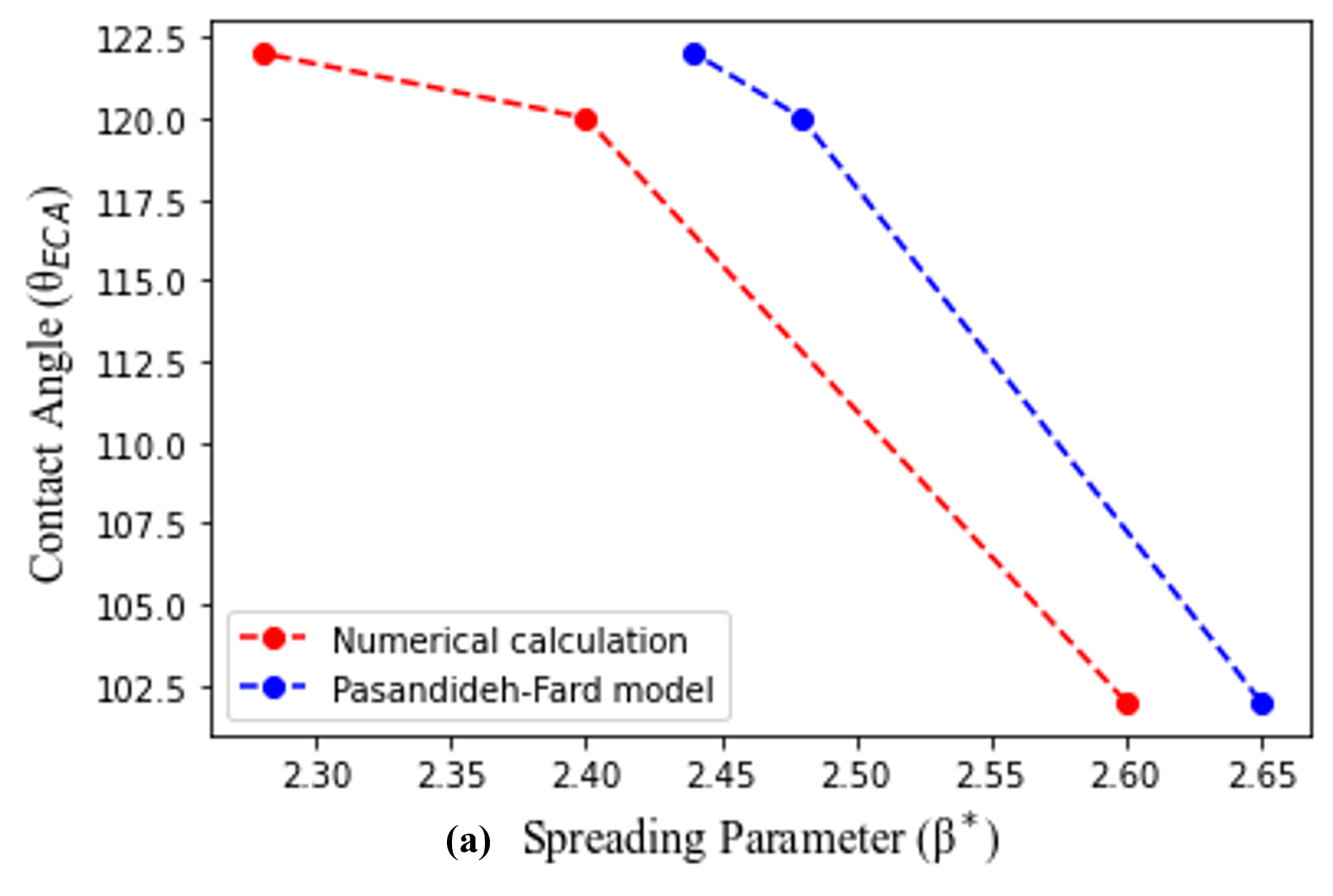} }}%
    %\qquad
    \subfloat{{\includegraphics[scale=0.71]{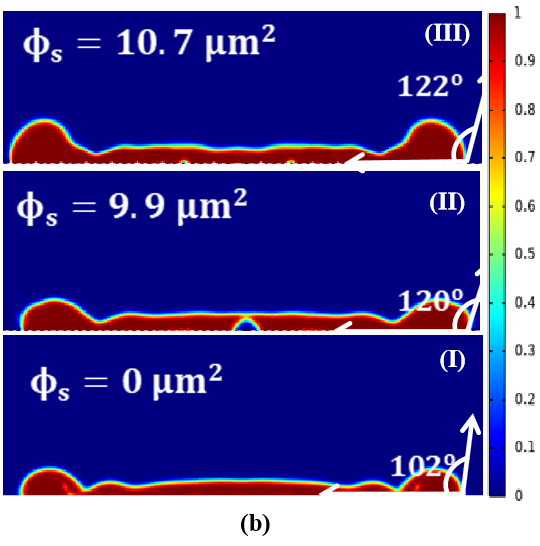} }}%
    \caption{(a) Comparison of spreading parameter ($\beta^*$) of the current study with \citep{doi:10.1063/1.868850} and (b) measured equilibrium contact $\theta_{ECA}$ with area fraction ($\phi_s$) covered by water on (I) plain, (II) triangular and (III) two-period reliefs surface.}%
    \label{fig:example}%
\end{figure}

%\begin{figure}
%    \centering
%    \subfigure{\includegraphics[scale=0.64]{np.png}} 
%    \subfigure{\includegraphics[scale=0.71]{fm.png}} 
%    %\subfigure[]{\includegraphics[width=0.24\textwidth]{monalisa.jpg}}
%    %\subfigure[]{\includegraphics[width=0.24\textwidth]{monalisa.jpg}}
%    \caption{(a) Comparison of spreading parameter ($\beta^*$) of the current study with \citep{doi:10.1063/1.868850} and (b) measured equilibrium contact $\theta_E_C_A$ with area fraction ($\phi_s$) covered by water on (I) plain, (II) triangular and (III) hierarchical surface reliefs. }
%    \label{fig:foobar}
%\end{figure}

\subsection{Effects of surface reliefs on maximum spreading based on energy analysis of droplet on several textured Ti-6Al-4V surfaces}
Furthermore, we compared the results of the maximum spreading diameter ($\beta^*$) of water droplet on the three surfaces with already established analytical models available in the literature and based on Re, We and $\theta_{ECA}$. Such scaling parameters and several scaling laws were extensively used to predict the maximum spreading diameter of liquid droplets on flat surfaces \citep{LIN201886}.  The applicability, consistency and reliability to capture droplet impact process of such laws are, however, very questionable \citep{LIN201886}. Here, we used these parameters to simplify the analysis, but not the laws themselves. In fact, because our model describes the droplet dynamics, we can also use it to verify the validity of these descriptions. 

\begin{equation}\label{equation_one}
\beta_{max} = \frac{d_{max}}{D_o} = \sqrt{\frac{We + 12}{3(1-cos(\theta_{ECA}))+4(We/\sqrt{Re})}}
\end{equation}

For this purpose, a simplified energy conservation model for a droplet on flat surfaces can be used. Among the available models, the analytical model of  Pasandideh-Fard et. al. \citep{doi:10.1063/1.868850}, which predicts the maximum $\beta^*$ after droplet impact based on energy conservation looks particularly interesting. This model uses the kinetic energy, surface energy and viscous dissipation before and after impact. The solid surface energy is assumed to be zero, that is, the total surface energy consists of only the liquid-vapor surface energy which is the product of the droplet surface area and the liquid-vapor surface tension \citep{https://doi.org/10.1002/aic.690430903}\citep{doi:10.1063/1.5006439}. The results of our numerical calculations and that of the analytical model for maximum $\beta_{max}$ (eqs. 10) of water droplet on plan Ti-6Al-4V and surfaces with triangular and two-period reliefs based on previously stated values of Re and We are summarised in Table 2. The maximum spreading diameter for flat Ti-6Al-4V surface of our calculation is in a good agreement with that of \citep{doi:10.1063/1.868850}. It is also interesting that the model predicts maximum $\beta^*$ for both triangular and a two-period reliefs Ti-6Al-4V surface. While the Pasandideh-Fard \citep{doi:10.1063/1.868850} model overestimates the maximum $\beta^*$ measured in experiment by 15\%, our numerical results deviate by 9.7\% from the results of this model. Fig. 7(a) shows such deviation in $\beta^*$. Additionally, Fig. 7 (bI, bII and bIII) shows the $\theta_{ECA}$ used in the calculations, and the fraction area ($\phi_s$) of the structure covered by water.

\begin{table}[ht]
\caption{Maximum $\beta^*$ of the current study and that of the analytical solution} % title of Table
\centering % used for centering table
\begin{tabular}{c c c c} % centered columns (4 columns)
\hline\hline %inserts double horizontal lines
Surfaces & Max. $\beta^*$ (current study) & Analytical solution \citep{doi:10.1063/1.868850} \\ [1ex] % inserts table
%heading
\hline % inserts single horizontal line
Plain Ti-6Al-4V  & 2.60 & 2.65 \\ % inserting body of the table
Triangular Ti-6Al-4V & 2.40 & 2.48 \\ [1ex] % [1ex] adds vertical space
Two-period reliefs Ti-6Al-4V & 2.28 & 2.44 \\ [1ex] % [1ex] adds vertical space
\hline %inserts single line
\end{tabular}
\label{table:nonlin} % is used to refer this table in the text
\end{table}

\subsection{Effect of velocity and viscosity of droplet on femtosecond textured Ti-6Al-4V relief surfaces}

\begin{figure}
    \centering
    \includegraphics[width=1\textwidth]{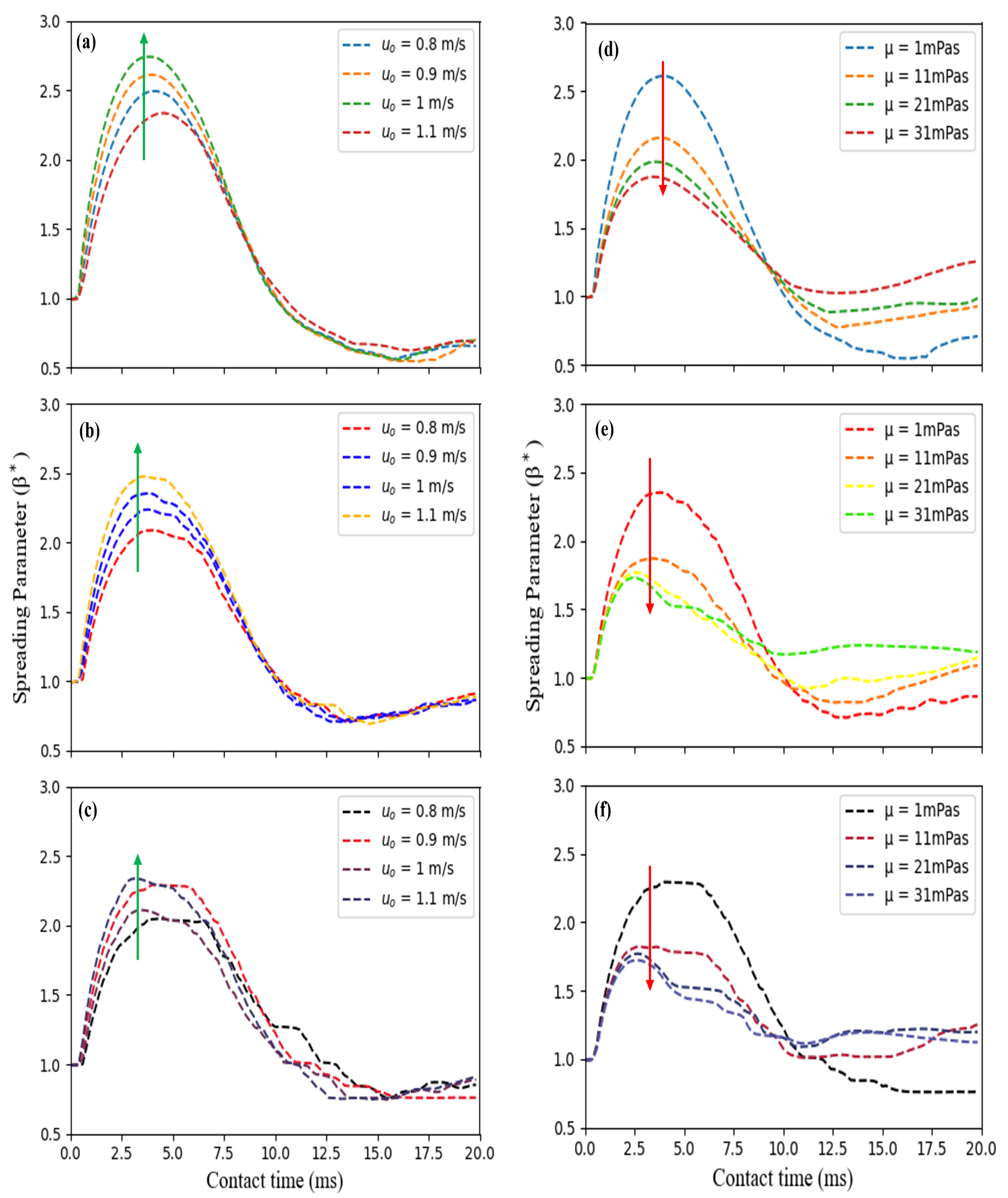}
    \caption{Spreading parameter ($\beta^*$) at $u_o$ = 0.8m/s - 1.1m/s (a) - (c) and $\mu$ = 1 mPas - 31 mPs (d) - (f) for flat, triangular and a two-period relief Ti-6Al-4V alloy surface.}
    \label{fig:First}
\end{figure}

%\begin{figure}
%    \centering
%    \subfigure{\includegraphics[scale=0.85]{vv.png}} 
%    %\subfigure{\includegraphics[scale=0.75]{Vis_New1.png}} 
%    %\subfigure[]{\includegraphics[width=0.24\textwidth]{monalisa.jpg}}
%    %\subfigure[]{\includegraphics[width=0.24\textwidth]{monalisa.jpg}}
%    \caption{Spreading parameter ($\beta^*$) at $u_o$ = 0.8m/s - 1.1m/s (a) - (f) and $\mu$ = 1 mPas - 31 mPs %for plan, triangular and hierarchical Ti-6Al-4V alloy surface.}
    \label{fig:foobar}
%\end{figure}

By varying the initial impact velocity ($u_o$) from 0.8 m/s to 1.1 m/s and the viscosity of droplet from 1 $\mu m$ to 31 $\mu m$ on flat, triangular and two-period reliefs Ti-6Al-4V surface, the evolution of the spreading parameter with the contact time are illustrated in Fig. 8 (a) - (f) respectively. For the simulation results shown in Fig. 8 (a) - (c), a dynamic viscosity of 1 mPas is set and the corresponding We is 21 - 39, while Re is 1876 - 2580. In addition, the velocity of  1 m/s is maintained for Fig. 8 (d) - (f), the We is 32 and Re ranges from 2345 to 77. The spreading parameter $\beta^*$ increases as the velocity rises for the three surfaces as presented in Fig 8(a) to (c). Interestingly, the contact time for the impact velocities, for which the droplet reaches a maximum $\beta^*$, is larger for a flat Ti-6Al-4V surface before the droplet starts to rapidly recoil on the surface. For triangular and two-period reliefs, the contact time is lower for the droplet to reach a maximum $\beta^*$ and this is due to the roughness scale. Though even when the velocity increases on the relief surfaces, droplet do not rapidly spread on the surfaces compared to that of flat Ti-6Al-4V surface. However, as the viscosity of the droplet increases, the impact process of the droplet on the surfaces is slowed down due to the viscous dissipation. As a result, the spreading parameter decreases as presented in Fig. 8 (d) to (f). The contact time, for which the droplet reaches a maximum $\beta^*$, becomes smaller on the triangular and two-period reliefs as a result of the roughness.

\subsection{Conclusion}

In this study, numerical studies of the effects of surface reliefs, impact velocity, and viscosity of liquid and droplet dynamics on  Ti-6Al-4V surfaces were examined. The main conclusions of this study are summarized as follows: 

(i) Firstly, on a flat Ti-6Al-4V surface, the droplet spreads rapidly due to the interplay between the inertial and capillary forces upon the impact with increased spreading parameter. For higher viscosity, the spreading process slows down with a decrease in the spreading parameter. The maximum spreading diameter of the droplet of the current study is in close agreement with the analytical model of Pasandideh-Fard \citep{doi:10.1063/1.868850} that is based on the energy balance for a liquid droplet on different surfaces.  

(ii) Small sub-micrometer surface reliefs affect the droplet speeding. In fact, the considered triangular relief with one period changes the wetting dynamics and increases the spreading parameter. Interestingly, the one with two periods and heights helps to furthermore confine the droplet, so that the droplet spreading is even more decreased. 

(iii) As impact velocity on these reliefs increases, the spreading diameter of the droplet rises.  The time it takes to reach the maximum spreading is, however, shorter than that on the flat Ti-6Al-4V surface. At low viscosity, the spreading parameter is high, but as the viscosity increases, the droplet diameter significantly decays.

Therefore, we conclude that triangular surface reliefs (or typical HSFL) alone are probably not well suitable for the increase in hydrophobic properties, but rather two-period (or more complex combinations of LSFL with HSFL of a femtosecond laser system) provide additional droplet capturing effects and are more promising for the wettability control.

\paragraph{\textbf{Acknowledgements:}}

This work was supported by FET Laser Implant  project (EU HORIZON 2020, Grant agreement ID: 951730) and the French Ministère de l'Éducation Nationale, de la Recherche et de la Technologie (MENRT) for the PhD scholarship of I.S.O.

\paragraph{\textbf{Conflicts of Interest:}} The authors declare no conflicts of interest.

\scriptsize
    {\bibliographystyle{elsarticle-num-names} 
        \bibliography{cas-refs}

\begin{thebibliography}{42}
\expandafter\ifx\csname natexlab\endcsname\relax\def\natexlab#1{#1}\fi
\providecommand{\url}[1]{\texttt{#1}}
\providecommand{\href}[2]{#2}
\providecommand{\path}[1]{#1}
\providecommand{\DOIprefix}{doi:}
\providecommand{\ArXivprefix}{arXiv:}
\providecommand{\URLprefix}{URL: }
\providecommand{\Pubmedprefix}{pmid:}
\providecommand{\doi}[1]{\href{http://dx.doi.org/#1}{\path{#1}}}
\providecommand{\Pubmed}[1]{\href{pmid:#1}{\path{#1}}}
\providecommand{\bibinfo}[2]{#2}
\ifx\xfnm\relax \def\xfnm[#1]{\unskip,\space#1}\fi
%Type = Article
\bibitem[{Bonse et~al.(2012)Bonse, Krüger, Höhm, and
  Rosenfeld}]{doi:10.2351/1.4712658}
\bibinfo{author}{J.~Bonse}, \bibinfo{author}{J.~Krüger},
  \bibinfo{author}{S.~Höhm}, \bibinfo{author}{A.~Rosenfeld},
\newblock \bibinfo{title}{Femtosecond laser-induced periodic surface
  structures},
\newblock \bibinfo{journal}{Journal of Laser Applications} \bibinfo{volume}{24}
  (\bibinfo{year}{2012}) \bibinfo{pages}{042006}.
%Type = Article
\bibitem[{Kietzig et~al.(2009)Kietzig, Hatzikiriakos, and
  Englezos}]{doi:10.1021/la8037582}
\bibinfo{author}{A.-M. Kietzig}, \bibinfo{author}{S.~G. Hatzikiriakos},
  \bibinfo{author}{P.~Englezos},
\newblock \bibinfo{title}{Patterned superhydrophobic metallic surfaces},
\newblock \bibinfo{journal}{Langmuir} \bibinfo{volume}{25}
  (\bibinfo{year}{2009}) \bibinfo{pages}{4821--4827}.
%Type = Article
\bibitem[{Bizi-bandoki et~al.(2013)Bizi-bandoki, Valette, Audouard, and
  Benayoun}]{BIZIBANDOKI2013399}
\bibinfo{author}{P.~Bizi-bandoki}, \bibinfo{author}{S.~Valette},
  \bibinfo{author}{E.~Audouard}, \bibinfo{author}{S.~Benayoun},
\newblock \bibinfo{title}{Time dependency of the hydrophilicity and
  hydrophobicity of metallic alloys subjected to femtosecond laser
  irradiations},
\newblock \bibinfo{journal}{Applied Surface Science} \bibinfo{volume}{273}
  (\bibinfo{year}{2013}) \bibinfo{pages}{399--407}.
%Type = Article
\bibitem[{Dahotre et~al.(2010)Dahotre, Paital, Samant, and
  Daniel}]{doi:10.1098/rsta.2010.0003}
\bibinfo{author}{N.~B. Dahotre}, \bibinfo{author}{S.~R. Paital},
  \bibinfo{author}{A.~N. Samant}, \bibinfo{author}{C.~Daniel},
\newblock \bibinfo{title}{Wetting behaviour of laser synthetic surface
  microtextures on ti–6al–4v for bioapplication},
\newblock \bibinfo{journal}{Philosophical Transactions of the Royal Society A:
  Mathematical, Physical and Engineering Sciences} \bibinfo{volume}{368}
  (\bibinfo{year}{2010}) \bibinfo{pages}{1863--1889}.
%Type = Article
\bibitem[{Liu et~al.(2004)Liu, Chu, and Ding}]{LIU200449}
\bibinfo{author}{X.~Liu}, \bibinfo{author}{P.~K. Chu},
  \bibinfo{author}{C.~Ding},
\newblock \bibinfo{title}{Surface modification of titanium, titanium alloys,
  and related materials for biomedical applications},
\newblock \bibinfo{journal}{Materials Science and Engineering; Reports}
  \bibinfo{volume}{47} (\bibinfo{year}{2004}) \bibinfo{pages}{49--121}.
%Type = Article
\bibitem[{Barthlott and Neinhuis(1997)}]{Barthlott1997}
\bibinfo{author}{W.~Barthlott}, \bibinfo{author}{C.~Neinhuis},
\newblock \bibinfo{title}{Purity of the sacred lotus, or escape from
  contamination in biological surfaces},
\newblock \bibinfo{journal}{Planta} \bibinfo{volume}{202}
  (\bibinfo{year}{1997}) \bibinfo{pages}{1--8}.
%Type = Article
\bibitem[{Vorobyev and Guo(2013)}]{https://doi.org/10.1002/lpor.201200017}
\bibinfo{author}{A.~Y. Vorobyev}, \bibinfo{author}{C.~Guo},
\newblock \bibinfo{title}{Direct femtosecond laser surface
  nano/microstructuring and its applications},
\newblock \bibinfo{journal}{Laser \& Photonics Reviews} \bibinfo{volume}{7}
  (\bibinfo{year}{2013}).
%Type = Article
\bibitem[{Wu et~al.(2021)Wu, He, Yin, Zhu, Xiao, Wu, and Duan}]{wu2021robust}
\bibinfo{author}{J.~Wu}, \bibinfo{author}{J.~He}, \bibinfo{author}{K.~Yin},
  \bibinfo{author}{Z.~Zhu}, \bibinfo{author}{S.~Xiao}, \bibinfo{author}{Z.~Wu},
  \bibinfo{author}{J.-A. Duan},
\newblock \bibinfo{title}{Robust hierarchical porous ptfe film fabricated via
  femtosecond laser for self-cleaning passive cooling},
\newblock \bibinfo{journal}{Nano Letters}  (\bibinfo{year}{2021}).
%Type = Article
\bibitem[{Yong et~al.(2018)Yong, Chen, Yang, Jiang, and
  Hou}]{doi.org/10.1002/admi.201701370}
\bibinfo{author}{J.~Yong}, \bibinfo{author}{F.~Chen},
  \bibinfo{author}{Q.~Yang}, \bibinfo{author}{Z.~Jiang},
  \bibinfo{author}{X.~Hou},
\newblock \bibinfo{title}{A review of femtosecond-laser-induced underwater
  superoleophobic surfaces},
\newblock \bibinfo{journal}{Advanced Materials Interfaces} \bibinfo{volume}{5}
  (\bibinfo{year}{2018}) \bibinfo{pages}{1701370}.
%Type = Article
\bibitem[{Cunha et~al.(2013)Cunha, Serro, Oliveira, Almeida, Vilar, and
  Durrieu}]{CUNHA2013688}
\bibinfo{author}{A.~Cunha}, \bibinfo{author}{A.~P. Serro},
  \bibinfo{author}{V.~Oliveira}, \bibinfo{author}{A.~Almeida},
  \bibinfo{author}{R.~Vilar}, \bibinfo{author}{M.-C. Durrieu},
\newblock \bibinfo{title}{Wetting behaviour of femtosecond laser textured
  ti-6al-4v surfaces},
\newblock \bibinfo{journal}{Applied Surface Science} \bibinfo{volume}{265}
  (\bibinfo{year}{2013}) \bibinfo{pages}{688--696}.
%Type = Article
\bibitem[{Klos et~al.(2020)Klos, Sedao, Itina, Helfenstein-Didier, Donnet,
  Peyroche, Vico, Guignandon, and Dumas}]{nano10050864}
\bibinfo{author}{A.~Klos}, \bibinfo{author}{X.~Sedao}, \bibinfo{author}{T.~E.
  Itina}, \bibinfo{author}{C.~Helfenstein-Didier}, \bibinfo{author}{C.~Donnet},
  \bibinfo{author}{S.~Peyroche}, \bibinfo{author}{L.~Vico},
  \bibinfo{author}{A.~Guignandon}, \bibinfo{author}{V.~Dumas},
\newblock \bibinfo{title}{Ultrafast laser processing of nanostructured patterns
  for the control of cell adhesion and migration on titanium alloy},
\newblock \bibinfo{journal}{Nanomaterials} \bibinfo{volume}{10}
  (\bibinfo{year}{2020}).
%Type = Article
\bibitem[{Young(1832)}]{doi:10.1098/rspl.1800.0095}
\bibinfo{author}{T.~Young},
\newblock \bibinfo{title}{An essay on the cohesion of fluids},
\newblock \bibinfo{journal}{Abstracts of the Papers Printed in the
  Philosophical Transactions of the Royal Society of London}
  \bibinfo{volume}{1} (\bibinfo{year}{1832}) \bibinfo{pages}{171--172}.
%Type = Article
\bibitem[{Wenzel(1949)}]{doi:10.1021/j150474a015}
\bibinfo{author}{R.~N. Wenzel},
\newblock \bibinfo{title}{Surface roughness and contact angle.},
\newblock \bibinfo{journal}{The Journal of Physical and Colloid Chemistry}
  \bibinfo{volume}{53} (\bibinfo{year}{1949}) \bibinfo{pages}{1466--1467}.
%Type = Article
\bibitem[{Wenzel(1936)}]{doi:10.1021/ie50320a024}
\bibinfo{author}{R.~N. Wenzel},
\newblock \bibinfo{title}{Resistance of solid surfaces to wetting by water},
\newblock \bibinfo{journal}{Industrial \& Engineering Chemistry}
  \bibinfo{volume}{28} (\bibinfo{year}{1936}) \bibinfo{pages}{988--994}.
%Type = Article
\bibitem[{Cassie and Baxter(1944)}]{TF9444000546}
\bibinfo{author}{A.~B.~D. Cassie}, \bibinfo{author}{S.~Baxter},
\newblock \bibinfo{title}{Wettability of porous surfaces},
\newblock \bibinfo{journal}{Trans. Faraday Soc.} \bibinfo{volume}{40}
  (\bibinfo{year}{1944}) \bibinfo{pages}{546--551}.
%Type = Article
\bibitem[{Onda et~al.(1996)Onda, Shibuichi, Satoh, and
  Tsujii}]{doi:10.1021/la950418o}
\bibinfo{author}{T.~Onda}, \bibinfo{author}{S.~Shibuichi},
  \bibinfo{author}{N.~Satoh}, \bibinfo{author}{K.~Tsujii},
\newblock \bibinfo{title}{Super-water-repellent fractal surfaces},
\newblock \bibinfo{journal}{Langmuir} \bibinfo{volume}{12}
  (\bibinfo{year}{1996}) \bibinfo{pages}{2125--2127}.
%Type = Article
\bibitem[{Shibuichi et~al.(1996)Shibuichi, Onda, Satoh, and
  Tsujii}]{doi:10.1021/jp9616728}
\bibinfo{author}{S.~Shibuichi}, \bibinfo{author}{T.~Onda},
  \bibinfo{author}{N.~Satoh}, \bibinfo{author}{K.~Tsujii},
\newblock \bibinfo{title}{Super water-repellent surfaces resulting from fractal
  structure},
\newblock \bibinfo{journal}{The Journal of Physical Chemistry}
  \bibinfo{volume}{100} (\bibinfo{year}{1996}) \bibinfo{pages}{19512--19517}.
%Type = Article
\bibitem[{Drelich and Chibowski(2010)}]{doi:10.1021/la1039893}
\bibinfo{author}{J.~Drelich}, \bibinfo{author}{E.~Chibowski},
\newblock \bibinfo{title}{Superhydrophilic and superwetting surfaces:
  Definition and mechanisms of control},
\newblock \bibinfo{journal}{Langmuir} \bibinfo{volume}{26}
  (\bibinfo{year}{2010}) \bibinfo{pages}{18621--18623}.
%Type = Article
\bibitem[{Qu{\'{e}}r{\'{e}}(2005)}]{2005}
\bibinfo{author}{D.~Qu{\'{e}}r{\'{e}}},
\newblock \bibinfo{title}{Non-sticking drops} \bibinfo{volume}{68}
  (\bibinfo{year}{2005}) \bibinfo{pages}{2495--2532}.
%Type = Article
\bibitem[{Bico et~al.(2002)Bico, Thiele, and Quéré}]{BICO200241}
\bibinfo{author}{J.~B. Bico}, \bibinfo{author}{U.~Thiele},
  \bibinfo{author}{D.~Quéré},
\newblock \bibinfo{title}{Wetting of textured surfaces},
\newblock \bibinfo{journal}{Colloids and Surfaces A: Physicochemical and
  Engineering Aspects} \bibinfo{volume}{206} (\bibinfo{year}{2002})
  \bibinfo{pages}{41--46}.
%Type = Article
\bibitem[{McHale(2007)}]{doi:10.1021/la7011167}
\bibinfo{author}{G.~McHale},
\newblock \bibinfo{title}{Cassie and wenzel: Were they really so wrong?},
\newblock \bibinfo{journal}{Langmuir} \bibinfo{volume}{23}
  (\bibinfo{year}{2007}) \bibinfo{pages}{8200--8205}.
%Type = Article
\bibitem[{Sikalo et~al.(2005)Sikalo, Wilhelm, Roisman, Jakirlić, and
  Tropea}]{doi:10.1063/1.1928828}
\bibinfo{author}{S.~Sikalo}, \bibinfo{author}{H.-D. Wilhelm},
  \bibinfo{author}{I.~V. Roisman}, \bibinfo{author}{S.~Jakirlić},
  \bibinfo{author}{C.~Tropea},
\newblock \bibinfo{title}{Dynamic contact angle of spreading droplets:
  Experiments and simulations},
\newblock \bibinfo{journal}{Physics of Fluids} \bibinfo{volume}{17}
  (\bibinfo{year}{2005}) \bibinfo{pages}{062103}.
%Type = Article
\bibitem[{Grewal et~al.(2014)Grewal, Cho, Oh, and Yoon}]{C4NR04069D}
\bibinfo{author}{H.~Grewal}, \bibinfo{author}{I.-J. Cho},
  \bibinfo{author}{J.-E. Oh}, \bibinfo{author}{E.-S. Yoon},
\newblock \bibinfo{title}{Effect of topography on the wetting of nanoscale
  patterns: experimental and modeling studies},
\newblock \bibinfo{journal}{Nanoscale} \bibinfo{volume}{6}
  (\bibinfo{year}{2014}) \bibinfo{pages}{15321--15332}.
%Type = Article
\bibitem[{Chamakos et~al.(2016)Chamakos, Kavousanakis, Boudouvis, and
  Papathanasiou}]{doi:10.1063/1.4941577}
\bibinfo{author}{N.~T. Chamakos}, \bibinfo{author}{M.~E. Kavousanakis},
  \bibinfo{author}{A.~G. Boudouvis}, \bibinfo{author}{A.~G. Papathanasiou},
\newblock \bibinfo{title}{Droplet spreading on rough surfaces: Tackling the
  contact line boundary condition},
\newblock \bibinfo{journal}{Physics of Fluids} \bibinfo{volume}{28}
  (\bibinfo{year}{2016}) \bibinfo{pages}{022105}.
%Type = Article
\bibitem[{Yagub et~al.(2015)Yagub, Farhat, Kondaraju, and Singh}]{YAGUB2015402}
\bibinfo{author}{A.~Yagub}, \bibinfo{author}{H.~Farhat},
  \bibinfo{author}{S.~Kondaraju}, \bibinfo{author}{T.~Singh},
\newblock \bibinfo{title}{A lattice boltzmann model for substrates with
  regularly structured surface roughness},
\newblock \bibinfo{journal}{Journal of Computational Physics}
  \bibinfo{volume}{301} (\bibinfo{year}{2015}) \bibinfo{pages}{402--414}.
%Type = Article
\bibitem[{Shan and Chen(1993)}]{PhysRevE.47.1815}
\bibinfo{author}{X.~Shan}, \bibinfo{author}{H.~Chen},
\newblock \bibinfo{title}{Lattice boltzmann model for simulating flows with
  multiple phases and components},
\newblock \bibinfo{journal}{Phys. Rev. E} \bibinfo{volume}{47}
  (\bibinfo{year}{1993}) \bibinfo{pages}{1815--1819}.
%Type = Article
\bibitem[{Olsson and Kreiss(2005)}]{OLSSON2005225}
\bibinfo{author}{E.~Olsson}, \bibinfo{author}{G.~Kreiss},
\newblock \bibinfo{title}{A conservative level set method for two phase flow},
\newblock \bibinfo{journal}{Journal of Computational Physics}
  \bibinfo{volume}{210} (\bibinfo{year}{2005}) \bibinfo{pages}{225--246}.
%Type = Inproceedings
\bibitem[{Hu et~al.(2014)Hu, Jia, Wan, and Xiong}]{hu2014simulation}
\bibinfo{author}{J.~Hu}, \bibinfo{author}{R.~Jia}, \bibinfo{author}{K.-t. Wan},
  \bibinfo{author}{X.~Xiong},
\newblock \bibinfo{title}{Simulation of droplet impingement on a solid surface
  by the level set method},
\newblock in: \bibinfo{booktitle}{Proceedings of the COMSOL Conference},
  \bibinfo{year}{2014}, pp. \bibinfo{pages}{8--10}.
%Type = Article
\bibitem[{COMSOL(2021)}]{2018}
\bibinfo{author}{COMSOL},
\newblock \bibinfo{title}{Comsol multiphysics}  (\bibinfo{year}{2021})
  \bibinfo{pages}{1742}.
%Type = Article
\bibitem[{Tadmor(2004)}]{doi:10.1021/la049410h}
\bibinfo{author}{R.~Tadmor},
\newblock \bibinfo{title}{Line energy and the relation between advancing,
  receding, and young contact angles},
\newblock \bibinfo{journal}{Langmuir} \bibinfo{volume}{20}
  (\bibinfo{year}{2004}) \bibinfo{pages}{7659--7664}.
%Type = Article
\bibitem[{Lin et~al.(2018)Lin, Zhao, Zou, Guo, Wei, and Chen}]{LIN201886}
\bibinfo{author}{S.~Lin}, \bibinfo{author}{B.~Zhao}, \bibinfo{author}{S.~Zou},
  \bibinfo{author}{J.~Guo}, \bibinfo{author}{Z.~Wei},
  \bibinfo{author}{L.~Chen},
\newblock \bibinfo{title}{Impact of viscous droplets on different wettable
  surfaces: Impact phenomena, the maximum spreading factor, spreading time and
  post-impact oscillation},
\newblock \bibinfo{journal}{Journal of Colloid and Interface Science}
  \bibinfo{volume}{516} (\bibinfo{year}{2018}) \bibinfo{pages}{86--97}.
%Type = Article
\bibitem[{Yokoi et~al.(2009)Yokoi, Vadillo, Hinch, and
  Hutchings}]{doi:10.1063/1.3158468}
\bibinfo{author}{K.~Yokoi}, \bibinfo{author}{D.~Vadillo},
  \bibinfo{author}{J.~Hinch}, \bibinfo{author}{I.~Hutchings},
\newblock \bibinfo{title}{Numerical studies of the influence of the dynamic
  contact angle on a droplet impacting on a dry surface},
\newblock \bibinfo{journal}{Physics of Fluids} \bibinfo{volume}{21}
  (\bibinfo{year}{2009}) \bibinfo{pages}{072102}.
%Type = Article
\bibitem[{Mehdi-Nejad et~al.(2003)Mehdi-Nejad, Mostaghimi, and
  Chandra}]{doi:10.1063/1.1527044}
\bibinfo{author}{V.~Mehdi-Nejad}, \bibinfo{author}{J.~Mostaghimi},
  \bibinfo{author}{S.~Chandra},
\newblock \bibinfo{title}{Air bubble entrapment under an impacting droplet},
\newblock \bibinfo{journal}{Physics of Fluids} \bibinfo{volume}{15}
  (\bibinfo{year}{2003}) \bibinfo{pages}{173--183}.
%Type = Article
\bibitem[{Mao et~al.(1997)Mao, Kuhn, and
  Tran}]{https://doi.org/10.1002/aic.690430903}
\bibinfo{author}{T.~Mao}, \bibinfo{author}{D.~C.~S. Kuhn},
  \bibinfo{author}{H.~Tran},
\newblock \bibinfo{title}{Spread and rebound of liquid droplets upon impact on
  flat surfaces},
\newblock \bibinfo{journal}{AIChE Journal} \bibinfo{volume}{43}
  (\bibinfo{year}{1997}) \bibinfo{pages}{2169--2179}.
%Type = Article
\bibitem[{Fang et~al.(2021)Fang, Li, Zhang, Zhu, Zhang, Li, Pan, Huang, Yang,
  Zheng, Yan, Huang, Maisotsenko, and Vorobyev}]{nano11040899}
\bibinfo{author}{R.~Fang}, \bibinfo{author}{Z.~Li}, \bibinfo{author}{X.~Zhang},
  \bibinfo{author}{X.~Zhu}, \bibinfo{author}{H.~Zhang},
  \bibinfo{author}{J.~Li}, \bibinfo{author}{Z.~Pan},
  \bibinfo{author}{Z.~Huang}, \bibinfo{author}{C.~Yang},
  \bibinfo{author}{J.~Zheng}, \bibinfo{author}{W.~Yan},
  \bibinfo{author}{Y.~Huang}, \bibinfo{author}{V.~S. Maisotsenko},
  \bibinfo{author}{A.~Y. Vorobyev},
\newblock \bibinfo{title}{Spreading and drying dynamics of water drop on hot
  surface of superwicking ti-6al-4v alloy material fabricated by femtosecond
  laser},
\newblock \bibinfo{journal}{Nanomaterials} \bibinfo{volume}{11}
  (\bibinfo{year}{2021}).
%Type = Article
\bibitem[{Varlamova et~al.(2019)Varlamova, Reif, Stolz, Borcia, Borcia, and
  Bestehorn}]{varlamova2019wetting}
\bibinfo{author}{O.~Varlamova}, \bibinfo{author}{J.~Reif},
  \bibinfo{author}{M.~Stolz}, \bibinfo{author}{R.~Borcia},
  \bibinfo{author}{I.~D. Borcia}, \bibinfo{author}{M.~Bestehorn},
\newblock \bibinfo{title}{Wetting properties of lipss structured silicon
  surfaces},
\newblock \bibinfo{journal}{The European Physical Journal B}
  \bibinfo{volume}{92} (\bibinfo{year}{2019}) \bibinfo{pages}{1--8}.
%Type = Article
\bibitem[{Dominic et~al.(2021)Dominic, Bourquard, Reynaud, Weck, Colombier, and
  Garrelie}]{dominic2021insignificant}
\bibinfo{author}{P.~Dominic}, \bibinfo{author}{F.~Bourquard},
  \bibinfo{author}{S.~Reynaud}, \bibinfo{author}{A.~Weck},
  \bibinfo{author}{J.-P. Colombier}, \bibinfo{author}{F.~Garrelie},
\newblock \bibinfo{title}{On the insignificant role of the oxidation process on
  ultrafast high-spatial-frequency lipss formation on tungsten},
\newblock \bibinfo{journal}{Nanomaterials} \bibinfo{volume}{11}
  (\bibinfo{year}{2021}) \bibinfo{pages}{1069}.
%Type = Article
\bibitem[{Liu et~al.(2021)Liu, Zhang, and Li}]{liu2021femtosecond}
\bibinfo{author}{R.~Liu}, \bibinfo{author}{D.~Zhang}, \bibinfo{author}{Z.~Li},
\newblock \bibinfo{title}{Femtosecond laser induced simultaneous functional
  nanomaterial synthesis, in situ deposition and hierarchical lipss
  nanostructuring for tunable antireflectance and iridescence applications},
\newblock \bibinfo{journal}{Journal of Materials Science \& Technology}
  \bibinfo{volume}{89} (\bibinfo{year}{2021}) \bibinfo{pages}{179--185}.
%Type = Article
\bibitem[{Thoroddsen et~al.(2005)Thoroddsen, Etoh, Takehara, Ootsuka, and
  Hatsuki}]{thoroddsen2005air}
\bibinfo{author}{S.~Thoroddsen}, \bibinfo{author}{T.~Etoh},
  \bibinfo{author}{K.~Takehara}, \bibinfo{author}{N.~Ootsuka},
  \bibinfo{author}{Y.~Hatsuki},
\newblock \bibinfo{title}{The air bubble entrapped under a drop impacting on a
  solid surface},
\newblock \bibinfo{journal}{Journal of Fluid Mechanics} \bibinfo{volume}{545}
  (\bibinfo{year}{2005}) \bibinfo{pages}{203--212}.
%Type = Article
\bibitem[{Schneider et~al.(2012)Schneider, Rasband, and
  Eliceiri}]{Schneider2012}
\bibinfo{author}{C.~A. Schneider}, \bibinfo{author}{W.~S. Rasband},
  \bibinfo{author}{K.~W. Eliceiri},
\newblock \bibinfo{title}{Nih image to imagej: 25 years of image analysis},
\newblock \bibinfo{journal}{Nature Methods} \bibinfo{volume}{9}
  (\bibinfo{year}{2012}) \bibinfo{pages}{671--675}.
%Type = Article
\bibitem[{Pasandideh-Fard et~al.(1996)Pasandideh-Fard, Qiao, Chandra, and
  Mostaghimi}]{doi:10.1063/1.868850}
\bibinfo{author}{M.~Pasandideh-Fard}, \bibinfo{author}{Y.~M. Qiao},
  \bibinfo{author}{S.~Chandra}, \bibinfo{author}{J.~Mostaghimi},
\newblock \bibinfo{title}{Capillary effects during droplet impact on a solid
  surface},
\newblock \bibinfo{journal}{Physics of Fluids} \bibinfo{volume}{8}
  (\bibinfo{year}{1996}) \bibinfo{pages}{650--659}.
%Type = Article
\bibitem[{Huang and Chen(2018)}]{doi:10.1063/1.5006439}
\bibinfo{author}{H.-M. Huang}, \bibinfo{author}{X.-P. Chen},
\newblock \bibinfo{title}{Energetic analysis of dropt's maximum spreading on
  solid surface with low impact speed},
\newblock \bibinfo{journal}{Physics of Fluids} \bibinfo{volume}{30}
  (\bibinfo{year}{2018}) \bibinfo{pages}{022106}.

\end{thebibliography}
        
        }





% \end{thebibliography}
\end{document}